\documentclass[3p]{elsarticle}

\usepackage{lineno,hyperref}
\usepackage{pgfplots}
\usepackage{tikzscale}
\usepackage[ruled,lined]{algorithm2e}
\pgfplotsset{compat=1.8}
\modulolinenumbers[5]

\journal{Future Generation Computer Systems}

\begin{document}

\begin{frontmatter}

\title{Comment on ``AndrODet: An adaptive Android obfuscation detector''}

\author{Alireza Mohammadinodooshan}
\ead{alireza.mohammadinodooshan@liu.se}
\author{Ulf Karg\'en\corref{mycorrespondingauthor}}
\cortext[mycorrespondingauthor]{Corresponding author}
\ead{ulf.kargen@liu.se}
\author{Nahid Shahmehri}
\ead{nahid.shahmehri@liu.se}
\address{Department of Computer and Information Science, Link\"oping University, SE-58183 Link\"oping, Sweden}

\begin{abstract}
We have identified a methodological problem in the empirical evaluation of the string encryption detection capabilities of the AndrODet system described by Mirzaei et al.\ in the recent paper ``AndrODet: An adaptive Android obfuscation detector''. The accuracy of string encryption detection is evaluated using samples from the AMD and PraGuard malware datasets. However, the authors failed to account for the fact that many of the AMD samples are highly similar due to the fact that they come from the same malware family. This introduces a risk that a machine learning system trained on these samples could fail to learn a generalizable model for string encryption detection, and might instead learn to classify samples based on characteristics of each malware family. Our own evaluation strongly indicates that the reported high accuracy of AndrODet’s string encryption detection is indeed due to this phenomenon. When we evaluated AndrODet, we found that when we ensured that samples from the same family never appeared in both training and testing data, the accuracy dropped to around 50\%. Moreover, the PraGuard dataset is not suitable for evaluating a static string encryption detector such as AndrODet, since the particular obfuscation tool used to produce the dataset effectively makes it impossible to extract meaningful features of static strings in Android apps.
\end{abstract}

\begin{keyword}
AndrODet \sep
Android \sep
Malware \sep
Obfuscation \sep
String encryption \sep
Machine learning
\end{keyword}

\end{frontmatter}

\section{Introduction}

In their paper ``AndrODet: An adaptive Android obfuscation detector'' \cite{Mirzaei2019}, Mirzaei at al. present the modular machine learning system AndrODet, which is capable of detecting three types of obfuscation in Android apps using statically extracted features: string encryption, identifier renaming, and control-flow obfuscation. AndrODet uses the MOA \cite{bifet2010moa} framework to perform on-line learning. The authors also compare their on-line approach with batch learning. In this comment paper, we have only considered the string encryption (SE) detection capabilities of AndrODet. Therefore, we restrict the discussion to SE detection for the remainder of the paper.

The authors propose several features for classifying apps as string encrypted, such as the average string entropy, average string length, etc., and evaluate on-line and batch models based on these features. For on-line learning, a combination of the AMD \cite{Wei2017} and PraGuard \cite{Maiorca2015} datasets are used for training and evaluation, whereas for batch-learning the model was trained on the AMD dataset and evaluated on the PraGuard dataset. The AMD set consists of 24,553 malware samples labeled with the obfuscation methods used by each sample (if any). PraGuard was constructed by collecting samples from two malware databases and running them through the DexGuard \cite{DexGuard} obfuscation tool.

The evaluation by Mirzaei et al.\ shows that, using the features proposed in the paper, both batch learning and on-line learning approaches can achieve over 80\% accuracy in SE detection. However, upon closer examination, we discovered a significant methodological problem in the way their evaluation is performed. While the AMD dataset has almost 25,000 malware samples, the samples in the dataset all belong to one of 71 malware families. Therefore, many samples share unique characteristics due to the fact that they belong to the same family. Moreover, for most malware families, all samples are in the same class, i.e., either all samples use SE or none of the samples use SE. (Only 3,883 samples belong to a family that has both SE and non-SE members.) This introduces a risk of \emph{memorization}, i.e., that the classifier learns a set of unique signatures of \emph{malware families}, instead of a generalizable model for detecting SE. Also, during evaluation of their approach, Mirzaei et al.\ make no effort to organize the dataset in a way that prevents malware of the same family from appearing in both the training and testing data. Therefore, memorization would also risk artificially inflating the measured accuracy of the model. Our own evaluation strongly suggests that the high accuracy that they report is indeed due to memorization. We trained and evaluated models on the AMD dataset with the same configuration and feature set as Mirzaei et al., using both batch learning and on-line learning. In the first experiment, 100 training and testing subsets were created by randomly sampling apps from the AMD dataset. In the second experiment, we made sure that the datasets were split so that samples from the same family never appeared in both the training and testing subsets. In the first experiment, we achieved an average accuracy of 92\% and 89\% for the batch and on-line cases respectively, which is similar to the accuracy reported by Mirzaei et al.\ However, in the second experiment, wherein we eliminated the possibility of learning a model based on the malware family, the respective accuracies dropped to 50\% and 51\%.

The dramatic drop in accuracy was surprising to us at first, as the accuracy when evaluating the batch learning model on the PraGuard dataset was very high according to the paper, which in turn seems to imply that the model can generalize well across different datasets. Unfortunately, however, the good performance on the PraGuard dataset appears to be coincidental. Since the DexGuard tool that was used to generate the PraGuard dataset employs a SE method that is fundamentally different from what is used by most other obfuscators, this dataset is not well-suited to evaluate how well an SE detector generalizes. Furthermore, the PraGuard dataset is particularly ill-suited to evaluate the AndrODet system, since the DexGuard tool effectively makes it impossible to statically extract the features used by AndrODet from encrypted strings. We elaborate further on this issue in Section \ref{sec:PraGuard}.

\section{Background}

In this section, we first provide some technical background on Android apps, which is required to understand the following discussion. We then briefly describe the AndrODet system.

Android apps are distributed in the form of Android application packages (APKs), which can in turn contain several DEX-files. Each DEX file contains static data and bytecode of one or more classes. Of particular interest to our discussion here is the \emph{string sectio}n of a DEX file, where all identifiers (method names, etc.) and constant strings used by the contained classes are stored. Each unique constant string has one entry in the string section. Obfuscation tools that implement SE replace the plaintext constant strings stored in the string section of an app with encrypted or scrambled versions and insert special decryption logic into the program code to decrypt strings just prior to their use. (Only non-identifier strings are relevant in the context of SE, as identifiers must be statically resolvable. Therefore, identifiers are obfuscated through renaming.)

The SE detection component of AndrODet retrieves all (non-identifier) strings from the string section of all DEX-files in an APK and extracts several features from the string material. The features are then fed to a machine learning model that determines whether or not the app uses SE. The following features are computed by AndrODet from the strings in the string section:

\begin{itemize}
	\item Average entropy
	\item Average wordsize
	\item Average length
	\item Average number of '\texttt{=}' characters
	\item Average number of '\texttt{-}' characters
	\item Average number of '\texttt{/}' characters
	\item Average number of '\texttt{+}' characters
	\item Average number of repeated characters
\end{itemize}

On a side note, we believe that the authors’ use of the term “average wordsize” is misleading since, based on the published source code of AndrODet, it turns out that the “wordsize” does not refer to the average size of words in strings. Instead, it is the size of the underlying Python string object used to represent strings in their implementation. It is unclear to us why the authors chose to use this as a feature, since the size of a string object is not solely determined by the number of bytes required to represent a string, but also depends on, for example, the underlying Python implementation and the memory allocator used by the OS. Also, the size of a string object is not guaranteed to be the same for two strings with the same content, or to be consistent between runs.

AndrODet uses the MOA platform for on-line learning, in which samples are fed to the on-line learning system in a streaming fashion. The \emph{prequential} evaluation mode of MOA was used for evaluation: When a new sample arrives from the stream, it is classified by the current model and the result is recorded. The model is then immediately updated with the new sample, and the process repeats for the next sample. The final accuracy is computed as the average percentage of correct classifications.  The authors also train a model using batch learning for comparison. For batch learning, they use the ATM framework \cite{Swearingen2017} for automatic tuning of model hyperparameters.

\section{Empirical Verification of the Problem}

We have evaluated both the batch and on-line learning approaches described by Mirzaei et al., using the published AndrODet source code. We used all of the samples from the AMD dataset, except for 135 samples in which the strings in the string section could not be decoded. We created the datasets for training and testing in the following way: First, 100 training and testing sets were constructed by splitting the AMD dataset into two sets of equal size using completely random sampling. Next, we created another 100 pairs of training and testing sets, by repeating the above procedure, with the added constraint that samples from the same malware family should never appear in both training and testing data. These two collections of training and testing sets will be referred to as the \emph{random} and \emph{non-overlapping} sets, respectively, for the remainder of this section. The exact procedure for creating the non-overlapping sets is outlined in Algorithm \ref{alg:nonoverlapping}.

\begin{algorithm}
	$F$: set of all families\\
	$S$: set of all samples\\
	$T_r$: set of samples in the training set\\
	$T_e$: set of samples in the testing set\\
	s($f$): a function returning all the samples belonging to family $f$\\
	rand($s$): a function returning a random sample from set $s$
	\BlankLine
	\SetKwInOut{Input}{input}\SetKwInOut{Output}{output}
	\Input{$S,F$}
	
	$T_r = \emptyset$\\
	\While{$|T_r| \leq \frac{|S|}{2}$}{
		$f$ = rand($F$)\\
		$F = F - f$\\
		$Tr = Tr \cup s(f)$
	}
	$T_e = S - T_r$
	
	\Output{$Tr,Te$}
	
	\caption{Procedure for constructing training and testing sets without overlapping malware families.}
	\label{alg:nonoverlapping}
\end{algorithm}

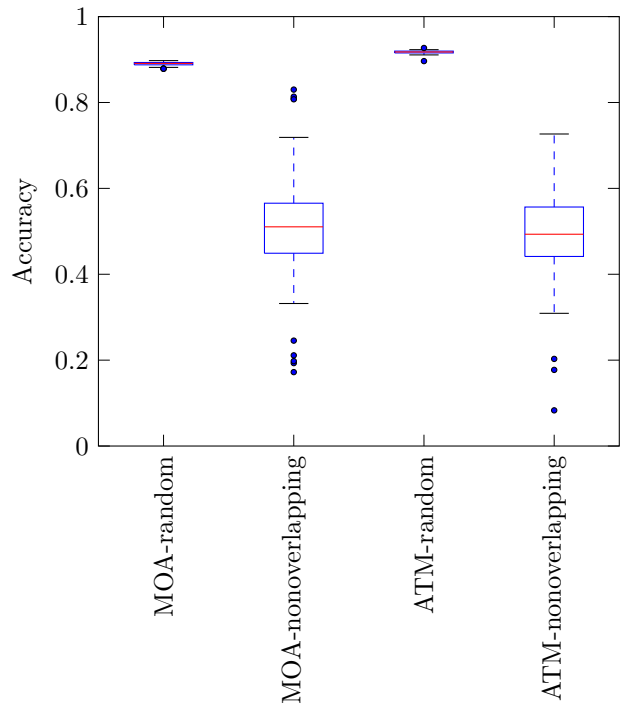
\begin{figure}
	\centering
\begin{tikzpicture}
\begin{axis}[
axis on top,
tick pos=both,
xmin=0.5, xmax=4.5,
xtick style={color=black},
xtick={1,2,3,4},
xticklabel style = {rotate=90.0},
xticklabels={MOA-random,MOA-nonoverlapping,ATM-random,ATM-nonoverlapping},
ylabel={Accuracy},
ymin=0, ymax=1,
ytick style={color=black}
]
\addplot [blue]
table {%
	0.775 0.88768531411254
	1.225 0.88768531411254
	1.225 0.893173069047424
	0.775 0.893173069047424
	0.775 0.88768531411254
};
\addplot [blue, dashed]
table {%
	1 0.88768531411254
	1 0.881726595134737
};
\addplot [blue, dashed]
table {%
	1 0.893173069047424
	1 0.897534605618806
};
\addplot [black]
table {%
	0.8875 0.881726595134737
	1.1125 0.881726595134737
};
\addplot [black]
table {%
	0.8875 0.897534605618806
	1.1125 0.897534605618806
};
\addplot [blue, mark=*, mark size=1, mark options={solid,draw=black}, only marks]
table {%
	1 0.879351298222623
	1 0.878532230321894
	1 0.878614137111967
};
\addplot [blue]
table {%
	1.775 0.448953868188499
	2.225 0.448953868188499
	2.225 0.565457452789083
	1.775 0.565457452789083
	1.775 0.448953868188499
};
\addplot [blue, dashed]
table {%
	2 0.448953868188499
	2 0.331892251413369
};
\addplot [blue, dashed]
table {%
	2 0.565457452789083
	2 0.718708718626156
};
\addplot [black]
table {%
	1.8875 0.331892251413369
	2.1125 0.331892251413369
};
\addplot [black]
table {%
	1.8875 0.718708718626156
	2.1125 0.718708718626156
};
\addplot [blue, mark=*, mark size=1, mark options={solid,draw=black}, only marks]
table {%
	2 0.193103448275862
	2 0.211149571871015
	2 0.19775890746739
	2 0.245456808563605
	2 0.172118618159584
	2 0.80905030757563
	2 0.813487972508591
	2 0.830046082949309
	2 0.807444061962134
};
\addplot [blue]
table {%
	2.775 0.915369809157179
	3.225 0.915369809157179
	3.225 0.919690392333524
	2.775 0.919690392333524
	2.775 0.915369809157179
};
\addplot [blue, dashed]
table {%
	3 0.915369809157179
	3 0.910803505610615
};
\addplot [blue, dashed]
table {%
	3 0.919690392333524
	3 0.923007617331477
};
\addplot [black]
table {%
	2.8875 0.910803505610615
	3.1125 0.910803505610615
};
\addplot [black]
table {%
	2.8875 0.923007617331477
	3.1125 0.923007617331477
};
\addplot [blue, mark=*, mark size=1, mark options={solid,draw=black}, only marks]
table {%
	3 0.896387910557785
	3 0.926939143254976
};
\addplot [blue]
table {%
	3.775 0.441470797594011
	4.225 0.441470797594011
	4.225 0.556582811685542
	3.775 0.556582811685542
	3.775 0.441470797594011
};
\addplot [blue, dashed]
table {%
	4 0.441470797594011
	4 0.309025650004377
};
\addplot [blue, dashed]
table {%
	4 0.556582811685542
	4 0.726636475916015
};
\addplot [black]
table {%
	3.8875 0.309025650004377
	4.1125 0.309025650004377
};
\addplot [black]
table {%
	3.8875 0.726636475916015
	4.1125 0.726636475916015
};
\addplot [blue, mark=*, mark size=1, mark options={solid,draw=black}, only marks]
table {%
	4 0.202896551724138
	4 0.0830752413918747
	4 0.177315805564048
};
\addplot [red]
table {%
	0.775 0.89102301580801
	1.225 0.89102301580801
};
\addplot [red]
table {%
	1.775 0.510450114148141
	2.225 0.510450114148141
};
\addplot [red]
table {%
	2.775 0.917642722581702
	3.225 0.917642722581702
};
\addplot [red]
table {%
	3.775 0.493219442035339
	4.225 0.493219442035339
};
\end{axis}
\end{tikzpicture}
	\caption{Box plot of classifier accuracy for the random and non-overlapping train/test configurations, using the batch (ATM) and on-line (MOA) machine learning frameworks.}
	\label{fig:boxplot-full}
\end{figure}

We repeated the experiments for each of the 100 train/test sets of the respective splitting strategies. Like Mirzaei et al., we used an SVM classifier for batch learning, and leveraging bagging for on-line learning. For the batch learning case we ran the ATM framework to find the best hyperparameters for each train/test set, using a budget of 200 configuration trials. For the on-line case, we first trained a model on the training set, and then used the model to classify the samples in the testing data (without updating the model). The results for batch and on-line learning are shown in Figure \ref{fig:boxplot-full}. For the on-line case, we get a mean accuracy of 89\% with the random sets, which is similar to the accuracy reported by Mirzaei et al.\ For the non-overlapping case, however, the accuracy drops to only 51\% on average (with a much greater variance across the 100 models). For the batch learning case we get similar results, with an average accuracy of 92\% for the random case, and 50\% for the non-overlapping case.

\begin{figure*}
	\includegraphics[width=\linewidth]{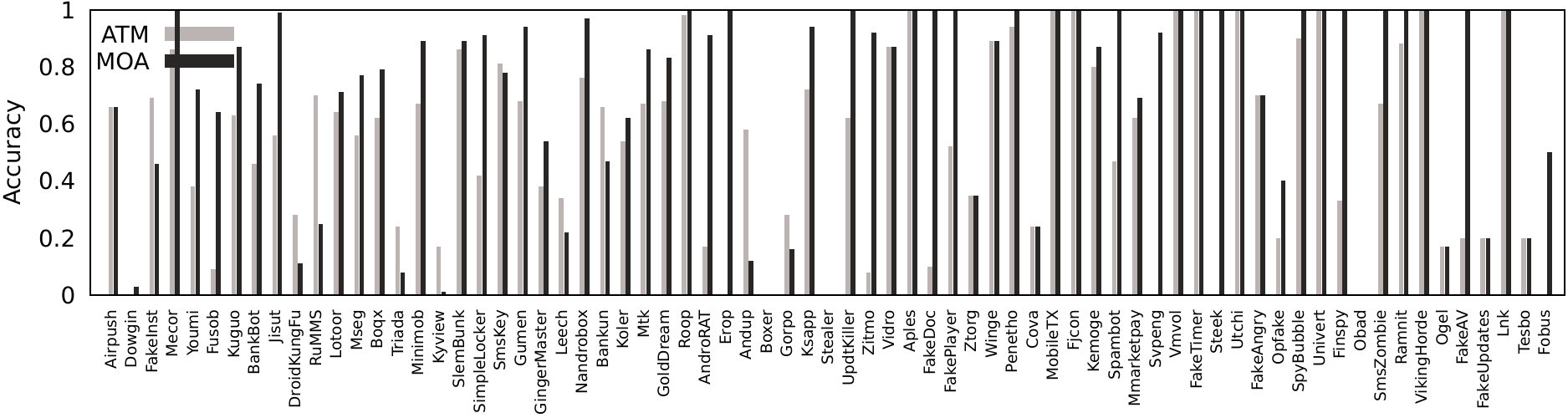}
	\caption{Individual classification accuracy of each family, when a classifier was trained on all other families in the AMD dataset.}
	\label{fig:acc-families}
\end{figure*}

One potential concern for us was that the distribution of samples from different families in the AMD dataset is highly skewed. For example, about one third of all samples belong to one family. This could, potentially, result in some training or testing sets containing samples from only a handful of families when we use our non-overlapping approach. To rule out the possibility that this was the cause of the reduced accuracy in the non-overlapping cases, we conducted another set of experiments where a classifier was trained on samples from all families except one, and the model was evaluated on the left-out family. We repeated this process for all families and computed the average accuracy, weighted by the number of samples in each family. The experiment was performed both for batch (ATM) and on-line learning (MOA), using the same parameters and setup as described above. Similarly to our first set of experiments, the average accuracy was 51\% for batch learning and 59\% for on-line learning. Moreover, the F-scores for the respective cases were 0.14 and 0.23. Figure \ref{fig:acc-families} shows the individual classification accuracy for each family. As can be seen from the figure, the accuracy varies widely between different families.

\section{Shortcomings of the PraGuard Dataset for Evaluating AndrODet}
\label{sec:PraGuard}

The PraGuard dataset was created by obfuscating all samples from the combined MalGenome and Contagio malware datasets using the obfuscation tool DexGuard. Mirzaei et al.\ used 1495 string-encrypted samples from PraGuard, along with the same number of non-obfuscated samples, to evaluate their batch-learned model. The reason that we believe PraGuard to be a poor choice for evaluating AndrODet is the particular way in which DexGuard implements SE. Most obfuscators that support SE work by substituting a plaintext string in the string section with an encrypted version. Therefore, the encrypted string can still be readily extracted from the string section and analyzed. However, DexGuard implements a more sophisticated form of SE that stores encrypted strings as byte arrays, rather than as actual strings in the string section. Since such byte arrays cannot be easily told apart from regular arrays used by the obfuscated app, DexGuard effectively prevents extraction and analysis of encrypted strings. Although DexGuard is closed source, independent analysis of DexGuard-obfuscated apps has confirmed this method of applying SE \cite{Moses2018}. Because DexGuard prevents AndrODet from extracting strings from the string section to compute features used for classification, we believe PraGuard is a poor choice for evaluating the generalizability of AndrODet. In fact, detecting DexGuard-obfuscated apps in which all (or almost all) constant strings have been removed from the string section is a trivial task that can be achieved with a simple heuristic, and does not require an advanced machine learning approach. We have analyzed the PraGuard dataset and confirmed that for 90\% of the SE samples \emph{all} non-identifier strings were removed from the string section\footnote{Additionally, for about 3\% of the samples, the AndrODet code for extracting strings failed to process the APK.}. The vast majority of the remaining samples had only a small number of non-identifier strings. We speculate that the relatively good performance on the PraGuard dataset reported by Mirzaei et al.\ is due to some DexGuard-encrypted samples appearing in the AMD dataset, allowing the classifier to learn to identify this particular type of SE. (For example, it would be trivial for a machine learning system to learn that an average string length of 0 indicates obfuscation.) However, as our experiments in the preceding section show, AndrODet fails to accurately detect SE in the general case.

\section{Discussion and Conclusion}

Our experimental evaluation of the SE detection capabilities of AndrODet strongly indicates that the system fails to learn a generalizable model for detecting SE, and instead bases its decision on the malware family to which a sample belongs. Since we obtained similar results with both the on-line and batch learning approaches proposed by Mirzaei et al., we believe the main problem to be the features used by AndrODet, rather than limitations of a specific machine learning approach. A potential explanation for the weak discriminative power of the features used to detect SE in AndrODet is that they are all based on compound statistics over all strings in an app. Since the percentage of strings that are actually subjected to encryption can vary dramatically between SE-obfuscated apps (i.e., it cannot be assumed that all strings in an SE-obfuscated app are encrypted) features such as the \emph{average} string entropy or length are not very informative. A more viable approach might be to detect SE on a per-string basis, and determine, for example, a threshold on the number of obfuscated strings for classifying an app as using SE.

Finally, it should be noted that we have only considered SE detection in this comment paper. Since we have not evaluated the performance of AndrODet’s detection of identifier renaming and control flow obfuscation, we can neither confirm nor rule out the possibility of similar problems when detecting these kinds of obfuscations. Since Mirzaei et al.\ use compound statistics for the entire app also when detecting these other two types of obfuscation, one potential concern is that AndrODet may fail to generalize to cases where only a subset of the identifiers, or only a fraction of the code of an app, has been obfuscated.

\smallskip
\noindent
\emph{In the interest of open science, we have made the code for running our experiments available at}\\ 
\begin{minipage}{\linewidth}
{\small \url{https://github.com/alirezamohammadinodooshan/androdet-se-eval}}
\end{minipage}

\section*{Acknowledgments}
The calculations were performed using computer resources provided by the Swedish National Infrastructure for Computing (SNIC) at the National Supercomputer Centre (NSC).

\bibliography{references}

\end{document}